\documentclass[letterpaper, paper,11pt]{AAS}		

\usepackage{bm}
\usepackage{amsmath}
\usepackage{subfigure}
\usepackage[colorlinks=true, pdfstartview=FitV, linkcolor=black, citecolor= black, urlcolor= black]{hyperref}
\usepackage{overcite}
\usepackage{footnpag}			      	
\usepackage{comment}
\usepackage{fancyhdr}
\PaperNumber{25-432}

\newcommand{\xtil}{\tilde{x}}
\newcommand{\ytil}{\tilde{y}}
\newcommand{\ztil}{\tilde{z}}
\newcommand{\pxtil}{\tilde{p}_x}
\newcommand{\pytil}{\tilde{p}_y}
\newcommand{\pztil}{\tilde{p}_z}

\newcommand{\FigFld}{./}

\renewcommand{\FigFld}{./Figures/}

\begin{document}

\title{Comparing Normal Form Representations for Station-Keeping near Cislunar Libration Points}
\author{Carson Hunsberger, \ David Schwab, \ 
Roshan Eapen,
\ and Puneet Singla 
}

\maketitle{}

\begin{abstract}
The normal forms provide useful approximations for many trajectories of interest within the circular restricted three-body problem. This paper aims to thoroughly compare two of these forms: the Birkhoff normal form and the resonant normal form, highlighting the strengths of each for the representation of center manifold trajectories. A method of station-keeping is introduced, analogous to Floquet modes, in which the unstable component is minimized at specific points along a trajectory through impulsive maneuvers. Three different formulations of the same station-keeping approach are posed, collectively spanning Lyapunov, vertical, and halo orbits, as well as Lissajous and quasihalo trajectories.

\end{abstract}

\section{Introduction}
There has been an increase in interest surrounding the circular restricted three-body problem (CR3BP) Hamiltonian normal forms over the last few years as their potential for elegant trajectory representation makes a strong case for their use in defining local orbital element equivalents near the libration points of the CR3BP. Lyapunov, vertical, and Lissajous trajectories-- along with their stable and unstable manifolds-- can all be parameterized in a straightforward manner using the action-angle variables associated with the Birkhoff normal form. The actions are first integrals of the normal form Hamiltonian, which is to say that they are constants, while the angles vary linearly with time. With a bit of massaging, the Birkhoff normal form is also able to parameterize a small section of the halo orbit family, which results from a bifurcation of the planar Lyapunov orbit. Unfortunately, quasihalo orbits cannot be parameterized using the Birkhoff action-angle variables. Following in the footsteps of Celletti\cite{cellettiLissajousHaloOrbits2015} and Peterson\cite{peterson2024toolkit}, certain terms can be kept in the Hamiltonian to produce the resonant normal form, which retains the quasihalo trajectories. A proper comparison between the trajectory parameterization capabilities of the Birkhoff and resonant normal forms has not yet been performed and is something that this paper aims to provide.

The second half of the paper focuses on formulating a normal form station-keeping approach, utilizing impulsive maneuvers, that can be used to stay near a desired periodic or quasiperiodic trajectory for an extended period of time. To account for the differences between the Birkhoff and resonant normal forms, three different variations of the station-keeping approach are posed. All three variations aim to minimize the unstable component in the normal form coordinates, denoted $\tilde{x}$, which defines the approximate distance of the spacecraft from a center manifold trajectory along the unstable manifold. In this way, the approach is similar to Floquet modes\cite{farres2022geometrical,gomezStationKeepingStrategiesTranslunar1998}.  Due to the constant nature of the Birkhoff actions, a single Newton iteration scheme can be formulated that minimizes the unstable coordinate $\tilde{x}$, as well as the distance from the nominal trajectory within the action-angle space. The remaining two variations deal with the resonant normal form. One tackles the quasiperiodic trajectories, in which only one action is explicitly minimized (in addition to $\tilde{x}$). Preliminary results show that by constraining only one action, the uncontrolled action will experience the desired fluctuation and the spacecraft will remain on the desired torus. Finally, a variation of the normal form station-keeping approach is introduced for halo orbits, which differs from the quasiperiodic case in that there is an additional constraint introduced to prevent the controlled trajectory from straying onto a nearby quasihalo trajectory.
\thispagestyle{fancy}
\fancyhf{}
\fancyfoot[C]{DISTRIBUTION A: Approved for public release; distribution is unlimited. Public Affairs release approval AFRL-2025-0080.}

\section{Birkhoff and Resonant Normal Forms of the CR3BP}
Before posing the normal form station-keeping formulation(s), a brief overview is provided of all of the transformations necessary for reducing the CR3BP Hamiltonian (about a libration point) to its Birkhoff normal form as well as its resonant normal form. For a more in-depth discussian and explanation of the Birkhoff normal form, refer to\cite{schwabCislunarTransportCharacterization2024,jorbaDynamicsCenterManifold1999,jorbaMethodologyNumericalComputation1999}. Further reading on the resonant normal form can be found in \cite{peterson2024toolkit,cellettiLissajousHaloOrbits2015}.

\subsection{Transformations}
Let the restricted three-body state be given by $\mathbf{x}_{RTB}=\left[x,y,z,\dot{x},\dot{y},\dot{z}\right]^T$. The restricted three-body state is first transformed to the local conjugate coordinates and momenta about a particular libration point, $\mathbf{x}_{\ell}=\left[X,Y,Z,P_X,P_Y,P_Z\right]^T$ by
\begin{equation}
    \mathbf{x}_{RTB} = VT\mathbf{x}_{\ell}+\mathbf{b},
\end{equation}
where
\begin{equation}
    V = \begin{bmatrix} 1&0&0&0&0&0\\0&1&0&0&0&0\\0&0&1&0&0&0\\0&1&0&1&0&0\\-1&0&0&0&1&0\\0&0&0&0&0&1 \end{bmatrix},\quad  T = \begin{bmatrix}\delta\gamma_j&0&0&0&0&0\\0&\delta\gamma_j&0&0&0&0\\0&0&\gamma_j&0&0&0\\0&0&0&\delta\gamma_j&0&0\\0&0&0&0&\delta\gamma_j&0\\0&0&0&0&0&\gamma_j \end{bmatrix},\quad \text{and} \quad \mathbf{b} = \begin{bmatrix} a - \mu\\0\\0\\0\\a-\mu\\0\end{bmatrix}.
\end{equation}
Here, $\gamma_j$ denotes the distance of the nearest primary from the libration point, and $\mu$ is the mass parameter of the CR3BP. For $L_1$ and $L_2$, $a=1\pm\gamma$ and $\delta=1$. In the case of the $L_3$ libration point, $a=-\gamma$ and $\delta=-1$. For details on the equilateral libration points, refer to \cite{jorbaMethodologyNumericalComputation1999}. This transformation translates the origin to the desired libration point, and scales the coordinates such that the distance from the new origin to the nearest primary is unitary.

The Hamiltonian of the CR3BP about a desired libration can then be expanded as a polynomial  in the elements of $\mathbf{x}_{\ell}$. For the $L_1$ libration point, the Hamiltonian takes the following form.
    \begin{equation}\label{Eq2}
    H=  \frac{1}{2}\left( P_X^2+P_Y^2+P_Z^2\right)+YP_X-XP_Y - c_2X^2 + \frac{c_2}{2}(Y^2+Z^2)-\sum_{n\geq 3}c_n\rho^nP_n\left(\frac{X}{\rho}\right)
\end{equation}
The coefficients $c_n$ result from the expansion of the potential energy about the libration point, and are discussed in greater depth in \cite{schwabCislunarTransportCharacterization2024}. From here, a linear, symplectic change of variables is applied which diagonalizes the quadratic part of the Hamiltonian.
\begin{equation}
\mathbf{x}_{qp}^{(2)}=C\mathbf{x}_{\ell}
\end{equation}
The process of finding $C$ is covered in detail in \cite{schwabCharacterizingAccuracyNormal2024, petersonOrbitalElementsRestricted2022, jorbaNumericalComputationNormal1998,angeljorbaLagrangianSolutions2015}. The quadratic part of the Hamiltonian, denoted $H_2$, can then be expressed in terms of the new state vector, $\mathbf{x}_{qp}^{(2)}=\left[x',y',z',p_z',p_y',p_z'\right]^T$
\begin{equation}
    H_2 = \lambda x' p_x' + \frac{\omega_1}{2}(y'^2 + p_y'^2) + \frac{\omega_2}{2}(z'^2 + p_z'^2)
\end{equation}
The diagonalized Hamiltonian can then be complexified with the transformation given by
\begin{equation}\label{Eq4}
    {}^{c}\mathbf{x}_{qp}^{(2)} = B\mathbf{x}_{qp}^{(2)},\quad B=\frac{1}{\sqrt{2}}\begin{bmatrix}\sqrt{2}&0&0&0&0&0\\ 0&1&0&0&i&0\\ 0&0&1&0&0&i\\ 0&0&0&\sqrt{2}&0&0\\ 0&i&0&0&1&0\\ 0&0&i&0&0&1 \end{bmatrix} 
\end{equation}
 The complexified coordinates can be denoted as ${}^{c}\mathbf{x}_{qp}^{(2)}=\left[q_1,q_2,q_3,p_1,p_2,p_3\right]^T$. The quadratic part of the Hamiltonian, expressed in the complexified variables, will then be
 \begin{equation}
     H_2 = \lambda q_1 p_1 + i\omega_2 q_2 p_2 + i \omega_2 q_3 p_3.
    \end{equation}
    The full Hamiltonian still has higher degree terms and can be expressed as the sum $H = H_2 + H_3 + H_4 + \dots$ where $H_n = \sum_{|\mathbf{k}_q|_1 + |\mathbf{k}_p|_1 = n} h_{\mathbf{k}_q\mathbf{k}_p}q_1^{k_{q1}}q_2^{k_{q2}}q_3^{k_{q3}}p_1^{k_{p1}}p_2^{k_{p2}}p_3^{k_{p3}}$. In practice, only the terms of degree $\leq N$ will be reduced to the normal form, with the rest considered a remainder, i.e., $H=H_2+\dots+H_N+R$. Through $N-2$ canonical Lie series transformations\cite{zhaoLieseriesTransformationsApplications2023b}, all of the undesirable terms of degree $3$ to degree $N$ are removed from the Hamiltonian. For the case of the Birkhoff normal form, all terms except those satisfying $\left<\mathbf{k}_p-\mathbf{k}_q,\boldsymbol{\zeta}\right>=0$ are removed, where $\boldsymbol{\zeta}=\left[\lambda,i\omega_2,i\omega_3\right]^T$. For the resonant normal form, all terms except those satisfying the more lenient condition, $k_{q_1} = k_{p_1}\,\text{and}\,(k_{p_2}-k_{q_2}) + (k_{p_3}-k_{q_3})=0$, are removed.

To remove the terms that do not satisfy the specified conditions (for either case), a series of generating functions, $G_n$, can be defined, that transform between the coordinates ${}^{c}\mathbf{x}_{qp}^{(n)}$ and ${}^{c}\mathbf{x}_{qp}^{(n-1)}$. Similar to $H_n$, $G_n = \sum_{|\mathbf{k}_q|_1 + |\mathbf{k}_p|_1 = n} g_{\mathbf{k}_q\mathbf{k}_p}q_1^{k_{q1}}q_2^{k_{q2}}q_3^{k_{q3}}p_1^{k_{p1}}p_2^{k_{p2}}p_3^{k_{p3}}$, where the coefficients $g_{\mathbf{k}_q\mathbf{k}_p}$ are given by
   \begin{equation}\label{eq:genfunc}
       g_{\mathbf{k}_q\mathbf{k}_p} = \frac{-h_{\mathbf{k}_q\mathbf{k}_p}}{\left<\mathbf{k}_p-\mathbf{k}_q,\boldsymbol{\zeta}\right>}.
   \end{equation}
   With the generating function $G_n$ defined, the Hamiltonian can be transformed through the following Lie series.
\begin{equation}
    H^{(n)} = H^{(n-1)} + \left\{H^{(n-1)},G_n\right\} +\frac{1}{2!} \left\{ \left\{ H^{(n-1)},G_n\right\}, G_n \right\} + \dots
\end{equation}
Note that this is the analytical equivalent of propagating the Hamiltonian according to the equations of motion defined by $G_n$ from time $t=0$ to $t=1$. The original complexified Hamiltonian would be denoted $H^{(2)}$ using this notation.
By applying the generating functions in a sequential manner ($G_3,\dots,G_N$), all of the undesirable terms can be removed from the Hamiltonian up to degree $N$. The coordinates and momenta will be similarly transformed by the generating functions, where
\begin{equation}
    p_k^{(n)} = p_k^{(n-1)} + \left\{p_k^{(n-1)},G_n\right\} + \frac{1}{2!}\left\{\left\{p_k^{(n-1)},G_n\right\},G_n\right\} + \dots
\end{equation}
provides the transformation from $\mathbf{x}_{qp}^{(n)}$ to the complex diagonal coordinates, $\mathbf{x}_{qp}^{(2)}$\cite{jorbaMethodologyNumericalComputation1999}. The inverse of this transformation is the following, 
\begin{equation}
    p_k^{(n-1)} = p_k^{(n)} + \left\{p_k^{(n)},-G_n\right\} + \frac{1}{2!}\left\{\left\{p_k^{(n)},-G_n\right\},-G_n\right\} + \dots ,
\end{equation}
which can be composed $N-2$ times to obtain the transformation from the complex diagonal coordinates, ${}^{c}\mathbf{x}_{qp}^{(2)}$, to the complex normal form coordinates, ${}^{c}\mathbf{x}_{qp}^{(N)}$.

The Hamiltonian is then cast back into real coordinates by performing the inverse of the complexification transformation, $B^{-1}:{}^{c}\mathbf{x}_{qp}^{(N)}\to \mathbf{x}_{NF}$, yielding the normal form state which is denoted as $\mathbf{x}_{NF}=\left[\tilde{x},\tilde{y},\tilde{z},\tilde{p}_x,\tilde{p}_y,\tilde{p}_z\right]^T$.
The normal form coordinates are then transformed into the Birkhoff action-angle coordinates, $\mathbf{x}_{AA}^B=\left[I_1,I_2,I_3,\phi_1,\phi_2,\phi_3\right]^T$, with $f_{AA}:\mathbf{x}_{NF}\to\mathbf{x}_{AA}^B$.
\begin{equation}
    \begin{matrix} \xtil = \sqrt{I_1}\exp{(\phi_1)} & \pxtil = \sqrt{I_1}\exp{(-\phi_1)} \\ \ytil = \sqrt{2I_2}\cos\phi_2 & \pytil = -\sqrt{2I_2}\sin\phi_2 \\ \ztil = \sqrt{2I_3}\cos\phi_3 & \pztil = -\sqrt{2I_3}\sin\phi_3 \end{matrix}
\end{equation}
After this step, the Birkhoff Hamiltonian is of the form $H^B = H^B(I_1,I_2,I_3)$ and has the corresponding equations of motion.
\begin{equation}
   \dot{I}_k = 0, \quad \dot{\phi}_k = \frac{\partial H}{\partial I_k} = const.
\end{equation}

Due to the presence of additional terms, the resonant Hamiltonian will instead have the form $H^R = H^R(I_1,I_2,I_3,\phi_2-\phi_3)$, meaning that the actions $I_2,I_3$ are no longer first integrals. Fortunately, their sum is a first integral, which leads to the final transformation for the reduction to the resonant normal form, $\mathbf{h}:\mathbf{x}_{AA}^B\to\mathbf{x}_{AA}^R$, where $\mathbf{x}_{AA}^R=\left[\hat{I}_1,\hat{I}_2,\hat{I}_3,\theta_1,\theta_2,\theta_3\right]^T$ denotes the resonant normal from action-angle coordinates\cite{peterson2024toolkit}.
\begin{equation}
    \begin{matrix}
   \hat{I}_1 = I_1 & \theta_1 = \phi_1 \\ \hat{I}_2 = I_2 & \theta_2 = \phi_2-\phi_3 \\ \hat{I}_3 = I_2+I_3 & \theta_3 = \phi_3 
    \end{matrix}
\end{equation}
This yields the resonant normal form Hamiltonian, $H^R = H^R(\hat{I}_1,\hat{I}_2,\hat{I}_3,\theta_2)$. More precisely, $H^R = H^R(\hat{I}_1,\hat{I}_2,\hat{I}_3,\cos\left(\theta_2\right))$. In general, $\hat{I}_2$ will not be a constant since the Hamiltonian depends explicitly on its conjugate coordinate, $\theta_2$. However, due to the nature of the terms in the Hamiltonian that contain $\theta_2$, there are three special cases that result in a constant $\hat{I}_2$: $\hat{I}_2=\hat{I}_3$, $\hat{I}_2=0$, or $\theta_2=\pm\frac{\pi}{2}$. These cases correspond to the Lyapunov family, the vertical family, and the halo family, respectively.
\subsection{Summary}
It is helpful to summarize all of the transformations necessary for reducing the CR3BP Hamiltonian to either of the normal forms. Let $\mathbf{g}: \mathbf{x}_{RTB} \to \mathbf{x}_{qp}^{(2)}$ denote the transformation from the state in the restricted three-body frame to the first set of real $q_i$ and $p_i$ coordinates. This transformation can be written as
\begin{equation}
    \mathbf{x}_{qp}^{(2)} = C^{-1}T^{-1}\left(V^{-1}\mathbf{x}_{RTB} - \mathbf{b}\right).
\end{equation}
At this point, two different approaches to the Lie series transformations can be taken. The first is referred to as the analytical transformation, $\mathcal{T}$, which is a set of six polynomials defining the final complex $\mathbf{x}_{qp}^{(N)}$ state in terms of the complex $\mathbf{x}_{qp}^{(2)}$ state. The normal form state, $\mathbf{x}_{NF}$, is reached upon the realification of $\mathbf{x}_{qp}^{(N)}$. From here, the transformation $f_{AA}:\mathbf{x}_{NF} \to \mathbf{x}_{AA}^B$ is applied to arrive at the Birkhoff action-angle coordinates. In the case of the resonant normal form, one additional transformation, $\mathbf{h}: \mathbf{x}_{AA}^B \to \mathbf{x}_{AA}^R$, must be applied. The full analytical transformations can be expressed as the following composition.
\begin{align}
    \mathbf{x}_{AA}^B &= \mathcal{A}_B\left(\mathbf{x}_{RTB}\right)=f_{AA}\circ B\circ \mathcal{T} \circ B^{-1}\circ\mathbf{g}(\mathbf{x}_{RTB})\\
    \mathbf{x}_{AA}^R &= \mathcal{A}_R\left(\mathbf{x}_{RTB}\right)=\mathbf{h}\circ f_{AA}\circ B\circ \mathcal{T} \circ B^{-1}\circ\mathbf{g}(\mathbf{x}_{RTB})
\end{align}
Alternatively, a numerical approach to the transformation can be taken. Rather than composing the $N-2$ Birkhoff transformations in polynomial form, one can simply numerically propagate the state according to the equations of motion prescribed by the generating function $G_n$ from $t=0$ to $t=-1$. This produces the near-identity transformation $\mathbf{x}_{qp}^{(n-1)}\to\mathbf{x}_{qp}^{(n)}$ (note that these are the realified $\mathbf{x}_{qp}$ coordinates), and by composing the sequence of all $N-2$ propagations, one arrives at $\mathcal{T}_{num}$. The full numerical transformation from the restricted three-body state to either of athe ction-angle states will then be the following.
\begin{align}
    \mathbf{x}_{AA}^B &= \mathcal{N}_B\left(\mathbf{x}_{RTB}\right)= f_{AA}\circ  \mathcal{T}_{num} \circ \mathbf{g}(\mathbf{x}_{RTB})\\
    \mathbf{x}_{AA}^R &= \mathcal{N}_R\left(\mathbf{x}_{RTB}\right)= \mathbf{h}\circ f_{AA}\circ  \mathcal{T}_{num} \circ \mathbf{g}(\mathbf{x}_{RTB})
\end{align}
A more in-depth explanation of the numerical transformation is provided in \cite{schwabCislunarTransportCharacterization2024}. All one needs to know is that the numerical transformation is generally more accurate for states within the region of validity. The trade-off is a greater computational cost, as $N-2$ initial value problems must be solved, rather than simply evaluating a polynomial, which is the case in the analytical transformation. Too far from the region of validity, however, the numerical transformation will fail to converge, while the analytical transformation will still provide an output (albeit an inaccurate one). The inverse of the analytical transformation, $\mathcal{A}^{-1}: \mathbf{x}_{AA}\to\mathbf{x}_{RTB}$, is more accurate than the inverse of the numerical transformation, $\mathcal{N}^{-1}$, and will therefore be used for all transformations from the action-angle coordinates to the restricted three-body coordinates.
\section{Representation of Center Manifold Trajectories}
One of the greatest strengths of the normal form, aside from having simplified dynamics, is the readily available parameterization of several common families of periodic orbits and quasiperiodic trajectories near the libration points. In certain cases, such as the horizontal Lyapunov family of orbits, the Birkhoff and resonant normal forms provide identical approximations, albeit with slightly different notation.
In going from the Birkhoff normal form to the resonant normal form, a first integral is lost in exchange for greater accuracy. The resulting increase in the region of validity allows for significantly better characterization of the vertical and halo families, as well as for Lissajous trajectories. Additionally, the resonant normal form is able to parameterize quasihalo trajectories, something that the Birkhoff normal form is incapable of due to its constant actions.
The remainder of this section covers the different parameterizations of each of the aforementioned trajectory families and provides a comparison between the representation capabilities of the two normal forms for trajectories on the center manifold. The capability of the normal forms to represent the stable and unstable manifolds of all of these center manifold trajectories is discussed in\cite{schwabCislunarTransportCharacterization2024,zhaoLieseriesTransformationsApplications2023b}.
For the purposes of this paper, it suffices to view the $\tilde{x}$ and $\tilde{p}_x$ normal form coordinates as indications of a state's distance from the center manifold along the unstable and stable manifolds, respectively.
\newpage
\subsection{Periodic Orbits}
\subsubsection{Lyapunov Orbits}\hfill

\begin{figure}[htb!]
    \centering
\begin{minipage}{0.28\textwidth} 
    \centering
    \includegraphics[width=\textwidth]{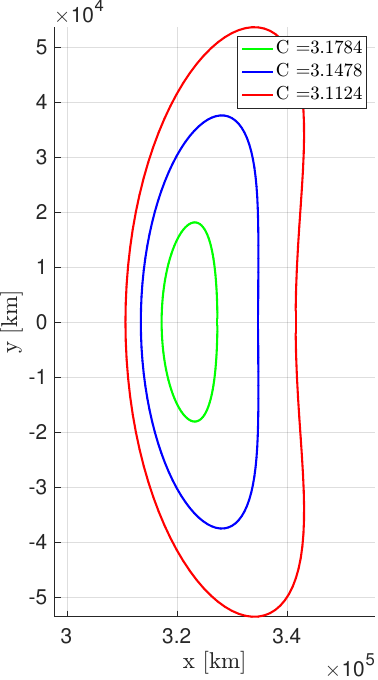} 
\end{minipage}%
\begin{minipage}{0.55\textwidth} 
    \centering
    \includegraphics[width=\linewidth]{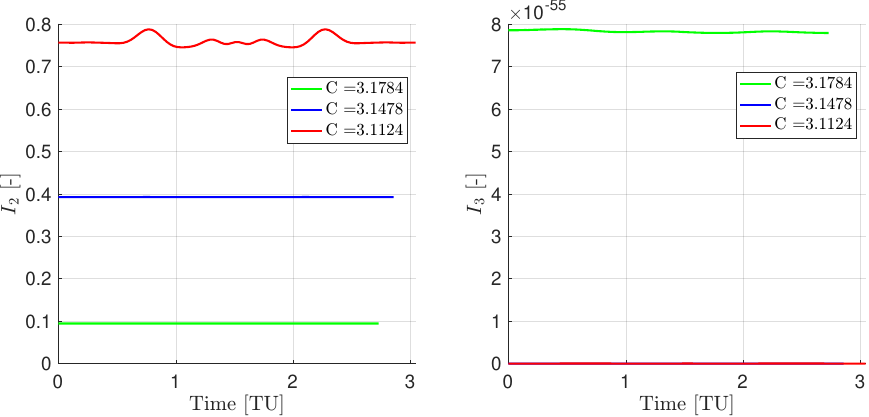} 
    \vspace{1em} 
    \includegraphics[width=\linewidth]{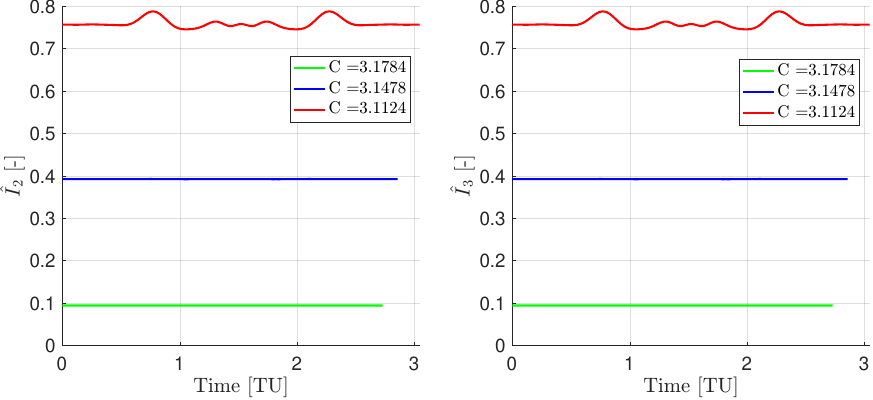} 
\end{minipage}
\caption{Comparison between the representation accuracy of the Birkhoff and resonant normal forms for numerical $L_1$ Lyapunov orbits, using the analytical transformation, $\mathcal{A}$.\label{EM-L1-Lyapunov}}
\end{figure}

Both the Birkhoff and resonant normal forms are able to represent a section of the horizontal Lyapunov family, albeit with slightly different notation.
In the Birkhoff action-angle coordinates, a Lyapunov orbit is parameterized by a nonzero, constant $I_2$ action, and $I_1=I_3=0$. The angle $\phi_2$ describes the location on the orbit and varies linearly with time.
\begin{equation}
\mathbf{x}^B_{AA} = \left[0,I_2,0,0,\phi_2(t),0  \right]\\
\end{equation}
The notation for a Lyapunov orbit in the resonant action-angle coordinates is slightly different due to the transformation $f_{AA}:\mathbf{x}_{AA}^B\to\mathbf{x}_{AA}^R$. This results in the actions $\hat{I}_2$ and $\hat{I}_3$ being both constant and equal, and the location on the orbit will be given by $\theta_2$ which varies linearly with time.
\begin{equation}
    \mathbf{x}^R_{AA} = \left[ 0, \hat{I}_2,\hat{I}_3,0,\theta_2(t),0 \right], \quad \hat{I}_2=\hat{I}_3
\end{equation}
To compare the representation capabilities of the two normal forms for the Lyapunov family of orbits, three orbits of increasing Jacobi constant are obtained numerically (through differential corrections). Along each of the orbits, one hundred $\mathbf{x}_{RTB}$ states are recorded and transformed into the Birkhoff and resonant action-angle coordinates using the analytical transformation $\mathcal{A}$. A constant action for the entire orbital period is the desired result, with fluctuation in the actions corresponding to poor parameterization. 
Figure \ref{EM-L1-Lyapunov} reveals that the resonant normal form is equivalent to the Birkhoff normal form for purely in-plane motion on the center manifold.

\subsubsection{Vertical Orbits}\hfill

Both normal forms can represent a section of the vertical family with just a single nonzero action, with a slight difference in notation due to the transformation $\mathbf{h}: \mathbf{x}_{AA}^B\to\mathbf{x}_{AA}^R$.
\begin{align}
    \mathbf{x}^B_{AA} &= \left[0,0,I_3,0,0,\phi_3(t)  \right]\\
    \mathbf{x}^R_{AA} &= \left[ 0, 0,\hat{I}_3,0,0,\theta_3(t) \right]
\end{align}

Note that while the transformation $\mathbf{h}$ would imply $\theta_2=\theta_3$ for the resonant case, the corresponding action $\hat{I}_2$ is equal to zero, meaning that any value of $\theta_2$ will map to the same $\mathbf{x}_{RTB}$ state, and so it can be set to zero for simplicity when working in the action-angle space.

\begin{figure}[htb!]
    \centering
\begin{minipage}{0.25\textwidth} 
    \centering
    \includegraphics[width=\textwidth]{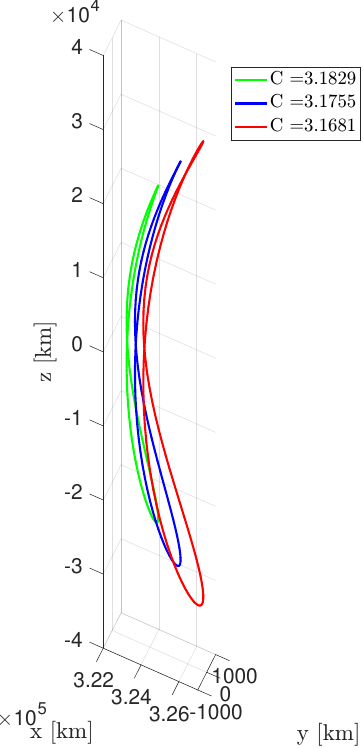} 
\end{minipage}%
\begin{minipage}{0.55\textwidth} 
    \centering
    \includegraphics[width=\linewidth]{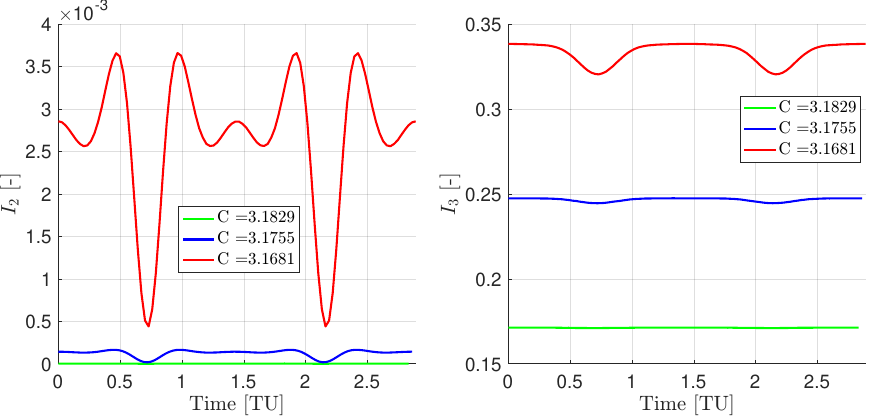} 
    \vspace{1em} 
    \includegraphics[width=\linewidth]{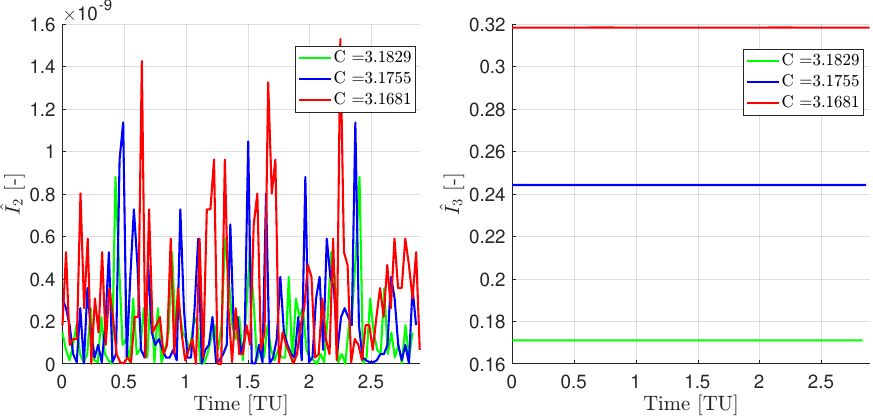} 
\end{minipage}
\caption{Comparison between the representation accuracy of the Birkhoff and resonant normal forms for numerical $L_1$ vertical orbits, using the analytical transformation, $\mathcal{A}$.\label{EM-L1-Vertical}}
\end{figure}

To test the ability of the normal forms to represent vertical orbits as the out-of-plane amplitude increases, three orbits about the $L_1$ libration point were considered in the same manner as the Lyapunov case. At 100 points along the orbits, the $\mathbf{x}_{RTB}$ state is transformed to the action-angler space with the analytical transformation $\mathcal{A}:\mathbf{x}_{RTB}\to\mathbf{x}_{AA}$. Figure \ref{EM-L1-Vertical} shows the three selected vertical orbits, as well as the center manifold actions in both the Birkhoff and resonant action-angle coordinates.

Unlike the parameterization of the Lyapunov orbits, in which the two normal forms are equivalent, the resonant normal form displays a significantly greater representation accuracy for vertical orbits. The Birkhoff approximation begins to fail at a Jacobi constant of $3.1755$, while further testing shows that the resonant action $\hat{I}_3$ remains constant up until a Jacobi constant of $3.13$ (corresponding to a $z$-amplitude of around $40000\,km$). This implies an increase in the region of validity for out-of-plane motion and has implications for the remaining center manifold trajectories that are yet to be discussed, all of which involve out-of-plane motion.
\subsubsection{Halo Orbits}\hfill

The representation of halo orbits using the normal forms is a bit more interesting than in the case of the Lyapunov or vertical families. In both the Birkhoff and the resonant action-angle coordinates, halo orbits are parameterized by two nonzero actions. Birkhoff normal form halo orbits require the additional constraint that the two angles, $\phi_2$ and $\phi_3$, vary linearly at identical rates and with a fixed separation of $\pm\frac{\pi}{2}\,rad$, where the sign denotes whether the orbit is in the northern or southern family. In practice, Birkhoff normal form halo orbits can be found by implementing a numerical method to minimize the constraint error, $\dot{\phi}_2-\dot{\phi}_3$, by varying one of the actions while fixing the other.
\begin{equation}
    \mathbf{x}^B_{AA} = \left[0,I_2,I_3,0,\phi_2(t),\phi_3(t)  \right],\quad \dot{\phi}_2 = \dot{\phi}_3\,\, \text{and}\,\,\phi_2 = \phi_3\pm\frac{\pi}{2}\,rad  \\
\end{equation}

In the resonant action-angle coordinates, this additional constraint is equivalent to fixing $\theta_2$ at a value of $\pm\frac{\pi}{2}\,rad$ while allowing $\theta_3$ to vary linearly with time. At the $L_1$ libration point, $\theta_2=\frac{\pi}{2}\, rad$ corresponds to the northern family while $\theta_2=-\frac{\pi}{2}\, rad$ corresponds to the southern family. At $L_2$, these signs flip. As mentioned earlier, $\theta_2$ taking either value results in the time derivative of $\hat{I}_2$ being zero. In order to find a halo orbit in the resonant action-angle coordinates, a numerical method can once again be implemented which minimizes $\dot{\theta}_2$ by varying $\hat{I}_2$ while keeping $\hat{I}_3$ fixed. 
\begin{equation}
    \mathbf{x}^R_{AA} = \left[ 0, \hat{I}_2,\hat{I}_3,0,\theta_2,\theta_3(t) \right],\quad \dot{\theta}_2 = 0\,\, \text{and}\,\, \theta_2 = \pm \frac{\pi}{2}\,rad
\end{equation}
\begin{figure}[htb!]
    \centering
\begin{minipage}{0.45\textwidth} 
    \centering
    \includegraphics[width=\textwidth]{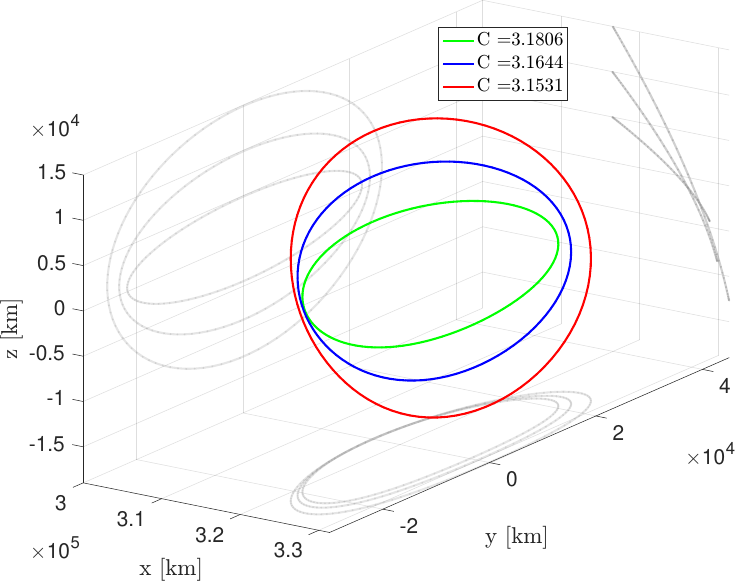} 
\end{minipage}%
\hfill
\begin{minipage}{0.55\textwidth} 
    \centering
    \includegraphics[width=\linewidth]{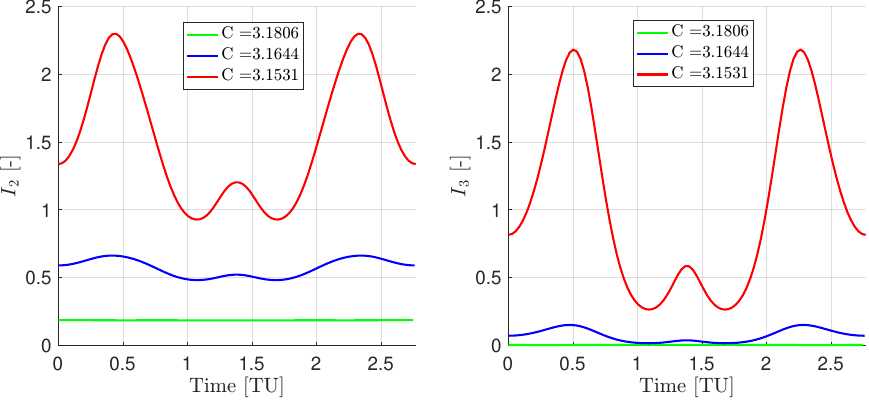} 
    \vspace{1em} 
    \includegraphics[width=\linewidth]{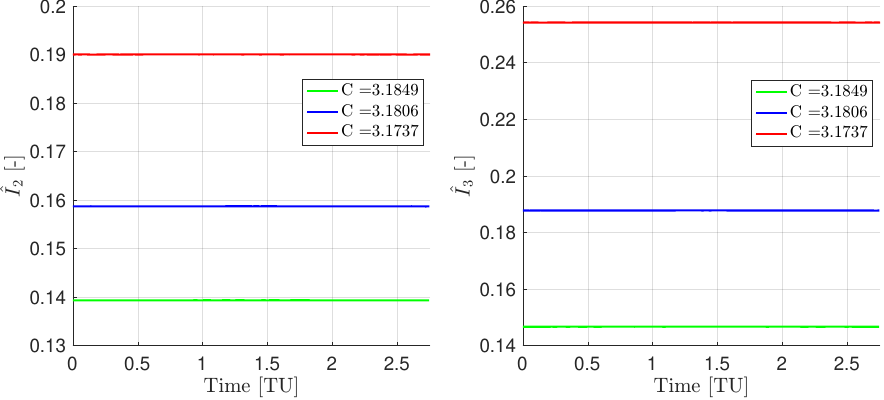} 
\end{minipage}
\caption{Comparison between the representation accuracy of the Birkhoff and resonant normal forms for small numerical $L_1$ halo orbits, using the analytical transformation, $\mathcal{A}$.\label{EM-L1-Halo}}
\end{figure}

To compare the halo orbit representation capabilities of the two normal forms, three northern $L_1$ halo orbits were selected, and at 100 points along each orbit, the instantaneous restricted three-body state was transformed into the two action-angle spaces using the analytical transformation, $\mathcal{A}$. Figure \ref{EM-L1-Halo} shows that the Birkhoff approximation of halo orbits is only valid when extremely close to the libration point. Upon further analysis, the resonant normal form actions begin to visibly fluctuate at a Jacobi constant of around $C=3.11$. 
\subsection{Quasiperiodic Trajectories}
\subsubsection{Lissajous Trajectories}\hfill

The only family of quasiperiodic trajectories that both of the normal forms are able to parameterize to some extent are the Lissajous trajectories. In the Birkhoff action-angle coordinates, these are simply parameterized by constant $I_2$ and $I_3$ actions, with the additional stipulation that the actions must not satisfy the constraints necessary for a halo orbit.
\begin{equation}
    \mathbf{x}^B_{AA} = \left[0,I_2,I_3,0,\phi_2(t),\phi_3(t)  \right],\quad \dot{\phi}_2 \neq \dot{\phi}_3
\end{equation}
The parameterization of a Lissajous trajectory in the resonant action-angle variables is substantially different from the Birkhoff case, as now $\hat{I}_2$ will be changing over time, and $\dot{\theta}_2$ will no longer be varying linearly with time. To validate that a particular trajectory is a Lissajous trajectory (rather than a quasihalo trajectory) while remaining within the action-angle space is to verify that the angle $\theta_2$ is always increasing.
\begin{equation}
\mathbf{x}^R_{AA} = \left[ 0, \hat{I}_2(t),\hat{I}_3,0,\theta_2(t),\theta_3(t) \right],\quad \dot{\theta}_2(t)>0\,\,\forall t 
\end{equation}
\begin{figure}[htb!]
    \centering
\begin{minipage}{0.3\textwidth} 
    \centering
    \includegraphics[width=\textwidth]{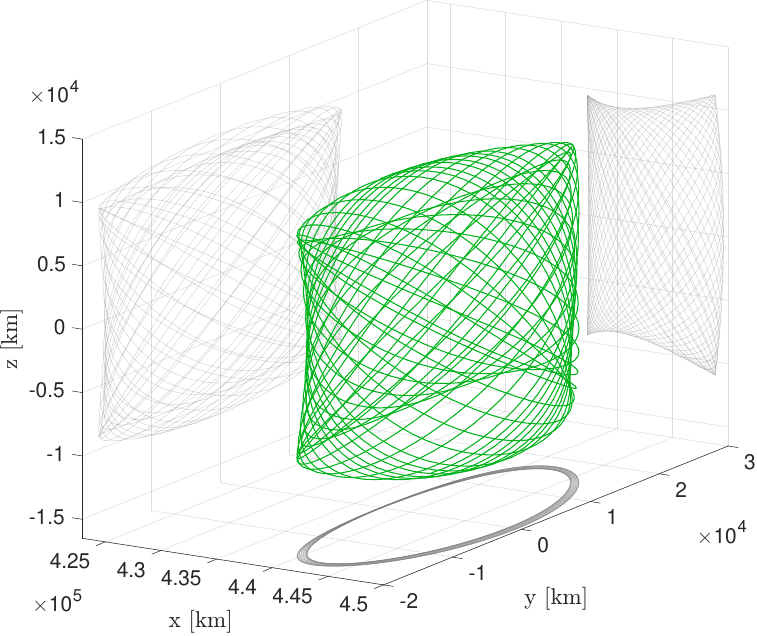} 
    \vspace{1em} 
    \includegraphics[width=\textwidth]{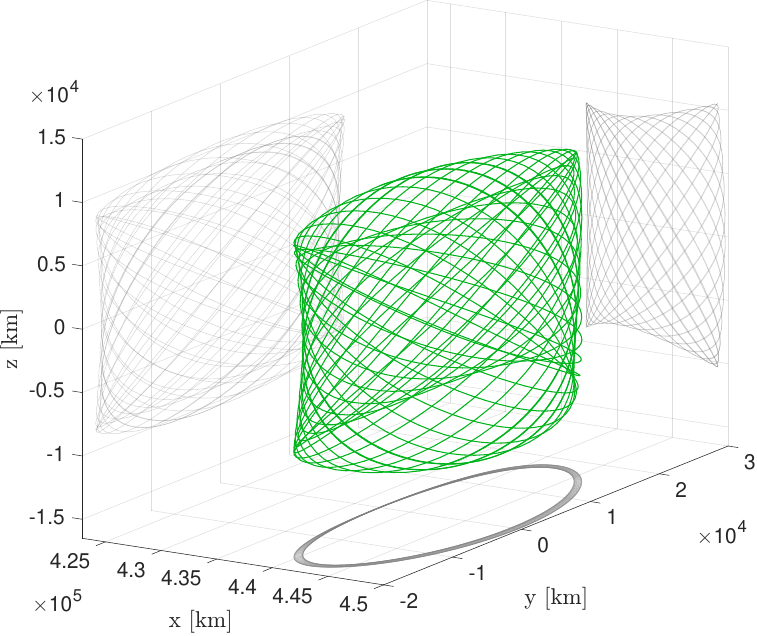}
    
\end{minipage}%
\begin{minipage}{0.55\textwidth} 
    \centering
    \includegraphics[width=\linewidth]{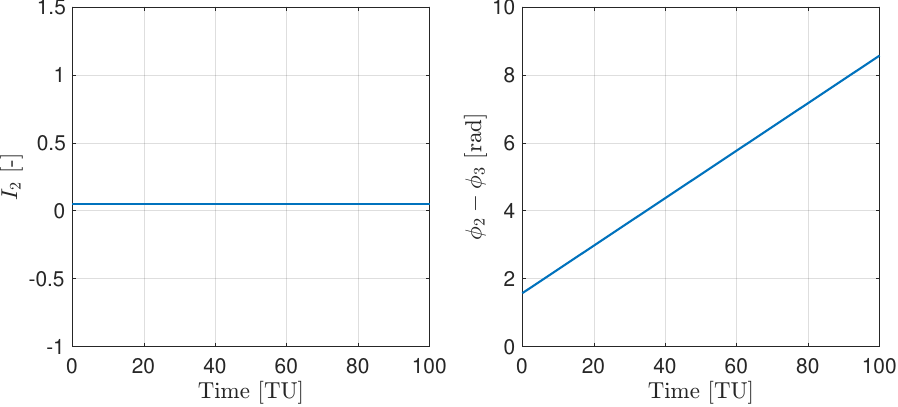}    
    \vspace{1em} 
    \includegraphics[width=\linewidth]{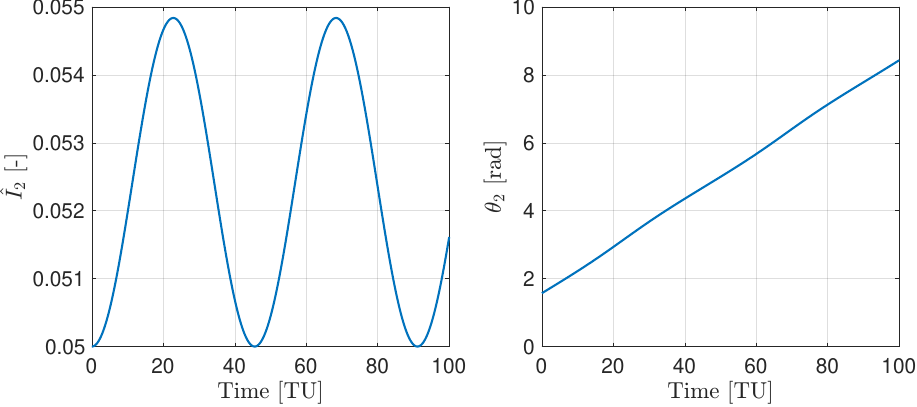} 
\end{minipage}
\caption{Comparison between two similar $L_2$ Lissajous trajectories obtained from the Birkhoff normal form (top) and the resonant normal form (bottom).\label{EM-L2-Lissajous}}
\end{figure}

Figure \ref{EM-L2-Lissajous} illustrates the key differences in the variation of the action-angle states for a Birkhoff and a resonant normal form Lissajous trajectory. An advantage of using the Birkhoff normal form is that it is trivial to propagate the state in action-angle space, however, the downside is a poorer approximation in the restricted three-body space. This difference in accuracy will be quantified later on in the station-keeping section.
\subsubsection{Quasihalo Trajectories}\hfill

The resonant normal form is the only approximation of the two that can represent quasihalo trajectories, due to the smaller region of validity of the Birkhoff normal form. In the resonant action-angle coordinates, a quasihalo trajectory is parameterized by two nonzero actions and is distinct from Lissajous trajectories because the angle $\theta_2$ will oscillate about $\theta_2^*=\pm\frac{\pi}{2}$, with the sign specifying the northern or southern family in the same way as for the halo orbits.
\begin{align*}
    \mathbf{x}^R_{AA} &= \left[ 0, \hat{I}_2(t),\hat{I}_3,0,\theta_2(t),\theta_3(t) \right],\quad |\theta_2(t)-\theta_2^*|<\frac{\pi}{2} \,\,\forall t 
\end{align*}
For an excellent visualization of the separation that exists between the Lissajous and quasihalo trajectories in both the action-angle and normal form spaces, refer to \cite{peterson2024toolkit}.

Since there is no comparison to be made between the two normal forms for quasihalo trajectories, the resonant normal form quasihalo trajectories are instead compared to their numerical counterparts. This is done using the Poincar\'e section approach \cite{kolemen2012multiple}, with the Jacobi constant and the area of the tori on the Poincar\'e section treated as constraints. Rather than beginning at the base halo orbit and continuing outward in area, which can be computationally expensive, the normal form provides feasible initial guesses to which the Newton iteration scheme can be applied directly, significantly decreasing the computation time necessary for finding a specific quasihalo torus.
\begin{figure}[htb!]
    \centering
    \begin{minipage}{\textwidth}
    \centering
    \includegraphics[width=\linewidth]{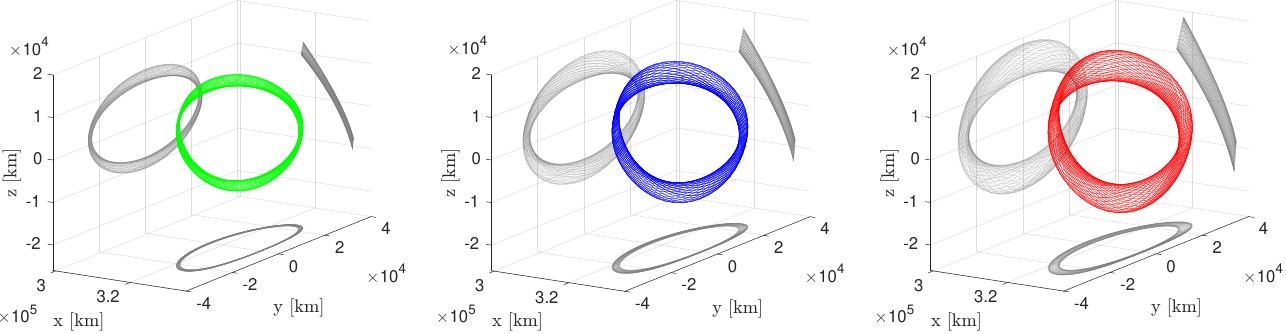}
    \end{minipage}
    \vspace{1em} 
    \begin{minipage}{0.6\textwidth}
    \centering
    \includegraphics[width=\linewidth]{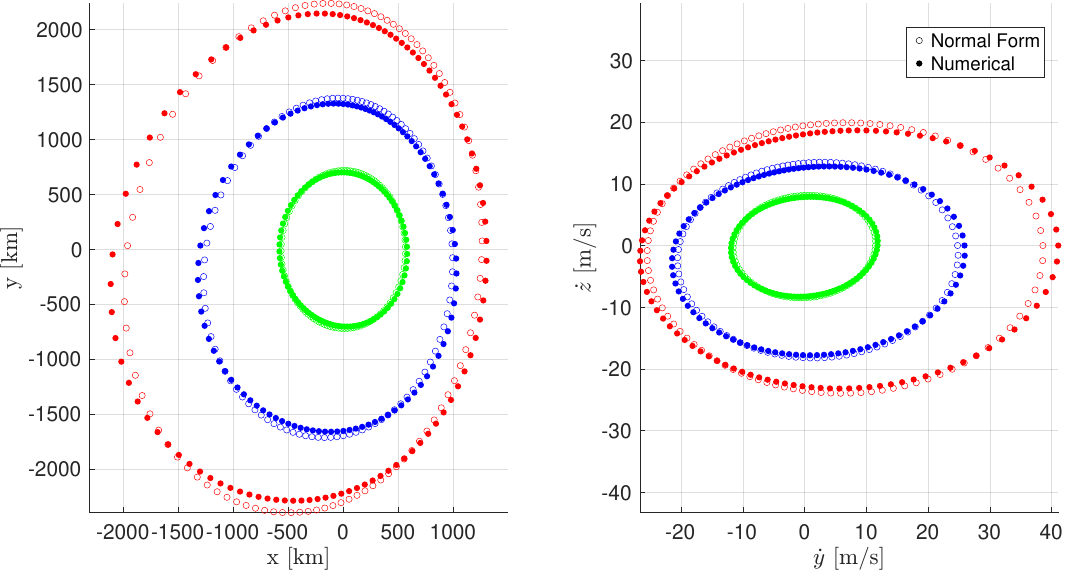}
    \end{minipage}
    \caption{Poincar\'e section comparison between three northern $L_1$ normal form quasihalo tori and their numerical counterparts.\label{ThreeQuasiHalo}}
\end{figure}

Figure \ref{ThreeQuasiHalo} depicts three normal form quasihalo tori of varying area, and plots two cross sections of their initial intersection with a particular Poincar\'e section (centered at a reference state), along with the numerical invariant curves obtained through Newton iteration. In all three cases, only a handful of iterations were necessary to converge to the numerical tori from the normal form tori. It is clear from Fig. \ref{ThreeQuasiHalo} that approximation error grows as the cross-sectional areas of the tori increase.

\clearpage

\section{Station-keeping}

With the parameterization capabilities of the normal forms in mind, one might seek a method of station-keeping that leverages either the Birkhoff or the resonant normal form. This paper proposes a straightforward implementation of Newton iteration as a means of identifying the impulsive maneuver, $\Delta\mathbf{v}$, that minimizes the spacecraft's distance from a desired trajectory in the action-angle space, at a given instant in time. The exact formulation varies slightly depending on which normal form is used, and which type of trajectory (periodic or quasiperiodic) is considered. Since the Birkhoff trajectories with nonzero out-of-plane motion are generally farther from the corresponding true numerical trajectories than the resonant normal form, it is expected (and verified in this section) that the utilization of the resonant normal form for station-keeping purposes will yield lower costs in terms of $\|\Delta\mathbf{v}\|$. As such, more analysis will be provided for the resonant normal form station-keeping scenarios.
\subsection{Birkhoff Station-keeping}
\subsubsection{Formulation} \hfill 

Let the restricted three-body state of the spacecraft be partitioned as $\mathbf{x}_{RTB}=\left[\mathbf{r},\mathbf{v}\right]^T$, where $\mathbf{v} = \left[\dot{x},\dot{y},\dot{z}\right]^T$. At a given instant in time, the position, $\mathbf{r}$, is fixed, while the velocity can be altered with an impulsive maneuver. Let the desired trajectory (which is assumed to lie on the center manifold) be parameterized by the actions $I_2^*$ and $I_3^*$ in the action-angle space. Using a combination of normal form coordinates and action-angle coordinates, the error vector $\mathbf{F}$ can be defined.
\begin{equation}
    \mathbf{F}(\mathbf{v}) =\begin{bmatrix}\tilde{x}\\ I_2-I_2^* \\ I_3-I_3^* \end{bmatrix} 
\end{equation}
Recall that $\tilde{x}$ coordinate indicates the distance from a nominal trajectory on the unstable manifold of the saddle subspace. It is clear that $\mathbf{F}(\mathbf{v}) = 0$ on only the desired trajectory and its stable manifold. Thus, by finding and applying a $\Delta\mathbf{v}$ that produces $\mathbf{F}(\mathbf{v}+\Delta\mathbf{v})=0$, the spacecraft will remain near the desired trajectory. If the initial state's velocity is denoted $\mathbf{v}_0$, the Newton iteration scheme in \eqref{Birkhoffnewton} can be leveraged to find the optimal $\Delta\mathbf{v}$.
\begin{equation}\label{Birkhoffnewton}
    \mathbf{v}_{k+1}=\mathbf{v}_k - \left(\left.\frac{d \mathbf{F}}{d \mathbf{v}}\right|_{\mathbf{v}_k}\right)^{-1}\mathbf{F}(\mathbf{v}_k)
\end{equation}
In practice, \eqref{Birkhoffnewton} should be iterated until a user-defined stopping criteria is reached. For the work done in this paper, the iterative process is ended when $\tilde{x}$ is brought below a value of $1\times10^{-14}$. Depending on the time interval between maneuvers, $\Delta t$, this can take anywhere from 2 to 6 iterations. When the iterative process ends, say after $m$ iterations, the optimal maneuver will be $\Delta\mathbf{v} = \mathbf{v}_m-\mathbf{v}_0$. By repeating the process of propagating the spacecraft's state in the restricted three-body frame for $\Delta t$ and then finding and applying the optimal $\Delta \mathbf{v}$, a long-term station-keeping methodology is born.

In \eqref{Birkhoffnewton}, the derivative $d\mathbf{F}/d\mathbf{v}$ will be the following $3\times 3$ matrix.
\begin{equation}\label{jac}
    \frac{d \mathbf{F}}{d \mathbf{v}} =\left[\begin{matrix}\frac{d \tilde{x}}{d \dot{x}}& \frac{d \tilde{x}}{d \dot{y}}& \frac{d \tilde{x}}{d \dot{z}}\\ \frac{d I_2}{d \dot{x}}& \frac{d I_2}{d \dot{y}}& \frac{d I_2}{d \dot{z}} \\ \frac{d I_3}{d \dot{x}}& \frac{d I_3}{d \dot{y}}& \frac{d I_3}{d \dot{z}} \end{matrix}\right] 
\end{equation}
It was determined that the analytical form of \eqref{jac}, found by simply differentiating the polynomials that make up the transformation $\mathcal{A}$, is insufficient for proper convergence of \eqref{Birkhoffnewton}, meaning that the derivatives must instead be obtained through the numerical transformation, $\mathcal{N}$.

\begin{equation}
    \frac{d \mathbf{x}_{NF}}{d \mathbf{x}_{RTB}} = \frac{d \mathbf{x}_{NF}}{d \mathbf{x}_{qp}^{(N-1)}}\frac{d \mathbf{x}_{qp}^{(N-1)}}{d \mathbf{x}_{qp}^{(N-2)}}\dots\frac{d \mathbf{x}_{qp}^{(3)}}{d \mathbf{x}_{qp}^{(2)}}\frac{d \mathbf{x}_{qp}^{(2)}}{d \mathbf{x}_{RTB}}
\end{equation}
Each derivative (except the rightmost one) is merely the state transition matrix (STM) of the transformation induced by a generating function, $-G$, at time $t=1$
\begin{equation}
    \frac{d \mathbf{x}_{qp}^{(n)}}{d \mathbf{x}_{qp}^{(n-1)}} \equiv \boldsymbol{\Phi}_n(1),
\end{equation}
where $\mathbf{x}_{NF} \equiv \mathbf{x}_{qp}^{(N)}$ since the numerical transformation uses the real $\mathbf{x}_{qp}$ coordinates. The STMs can be found by propagating \eqref{eq:stm} from $t=0$ to $t=1$
\begin{equation}\label{eq:stm}
    \dot{\boldsymbol{\Phi}}_n(t) = A_n(t)\boldsymbol{\Phi}_n(t),
\end{equation}
where $\boldsymbol{\Phi}(0) = I_{6\times 6}$ and $A_n(t)$ is given by

\begin{equation}
    A_n(t) = -J\left.\frac{\partial^2 G}{\partial \mathbf{x}_{qp}^2}\right|_{\mathbf{x}_{qp}(t)}, \quad\text{with} \quad J=\begin{bmatrix}0& I_{3\times 3} \\ -I_{3\times 3}& 0\end{bmatrix}.
\end{equation}

Upon multiplying all of the STMs together, one arrives at $\frac{d \mathbf{x}_{NF}}{d \mathbf{x}_{qp}^{(2)}}$. The final component is then
\begin{equation}
    \frac{d \mathbf{x}_{qp}^{(2)}}{d \mathbf{x}_{RTB}}=C^{-1}T^{-1}V^{-1},
\end{equation}
where the $6\times 6$ matrices $C$, $T$, and $V$ are the same as the ones mentioned earlier in the normal form reduction process. Note that due to the symplectic nature of $C$, the inverse can be written as $C^{-1} = -JC^TJ$. When working with higher-order approximations, it is beneficial to avoid the errors typically introduced through solving for the inverse in a numerical manner.

With the derivatives of the normal form coordinates with respect to the restricted three-body frame coordinates calculated, it is straightforward to obtain the first row of \eqref{jac}.
\begin{equation}
    \frac{d \tilde{x}}{d \dot{x}} = \left[  \frac{d \mathbf{x}_{NF}}{d \mathbf{x}_{RTB}} \right]_{1,4}\quad \frac{d \tilde{x}}{d \dot{y}}= \left[  \frac{d \mathbf{x}_{NF}}{d \mathbf{x}_{RTB}} \right]_{1,5}\quad  \frac{d \tilde{x}}{d \dot{z}} = \left[  \frac{d \mathbf{x}_{NF}}{d \mathbf{x}_{RTB}} \right]_{1,6}
\end{equation}
The remaining entries are obtained by taking into account the derivatives of $I_2$ and $I_3$ with respect to the restricted three-body state.
\begin{equation}
    \frac{d I_2}{d \mathbf{x}_{RTB}} = \frac{d I_2}{d \mathbf{x}_{NF}}\frac{\mathbf{x}_{NF}}{\mathbf{x}_{RTB}}, \quad  \frac{d I_3}{d \mathbf{x}_{RTB}} = \frac{d I_3}{d \mathbf{x}_{NF}}\frac{\mathbf{x}_{NF}}{\mathbf{x}_{RTB}}
\end{equation}
\begin{equation}
    \frac{d I_2}{d \mathbf{x}_{NF}} = \begin{bmatrix}0& \tilde{y} &0 &0 &\tilde{p}_y &0\end{bmatrix},\quad  \frac{d I_3}{d \mathbf{x}_{NF}} = \begin{bmatrix}0&0& \tilde{z} &0 &0 &\tilde{p}_z \end{bmatrix}
\end{equation}

This station-keeping formulation leveraging the Birkhoff normal form is quite flexible. In order to stay on a Lyapunov orbit, one merely needs to set $I_3^*=0$. For a vertical orbit, $I_2^*=0$. In the case of a Lissajous trajectory or a (small) halo orbit, both desired actions will be nonzero.  
\subsubsection{Results}\hfill

\begin{figure}[htb!]
    \centering
    \subfigure[\label{subfig:BirkhoffLissajous-RTB}]{
        {\includegraphics[width=0.6\textwidth]{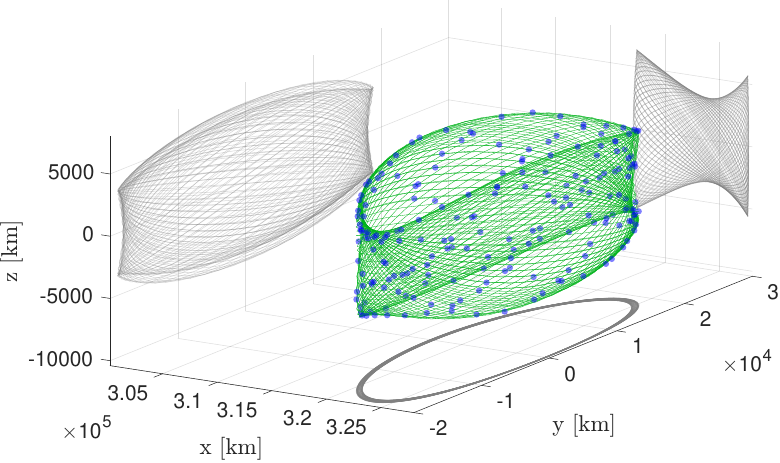}}
        }
    \subfigure[\label{subfig:BirkhoffLissajous-Control}]{
        {\includegraphics[width=0.35\textwidth]{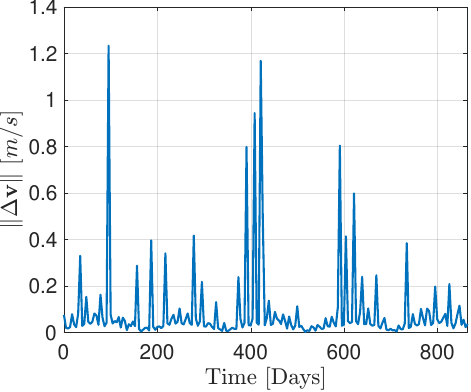}}
        }
        \caption{Station-keeping of an $L_1$ Birkhoff Lissajous trajectory ($t_f = 200\,TU$,\,\,$\Delta t = 1\,TU$).\label{fig:BirkhoffLissajous}}
\end{figure}

\begin{figure}[htb!]
    \centering
    \includegraphics[width=1\textwidth]{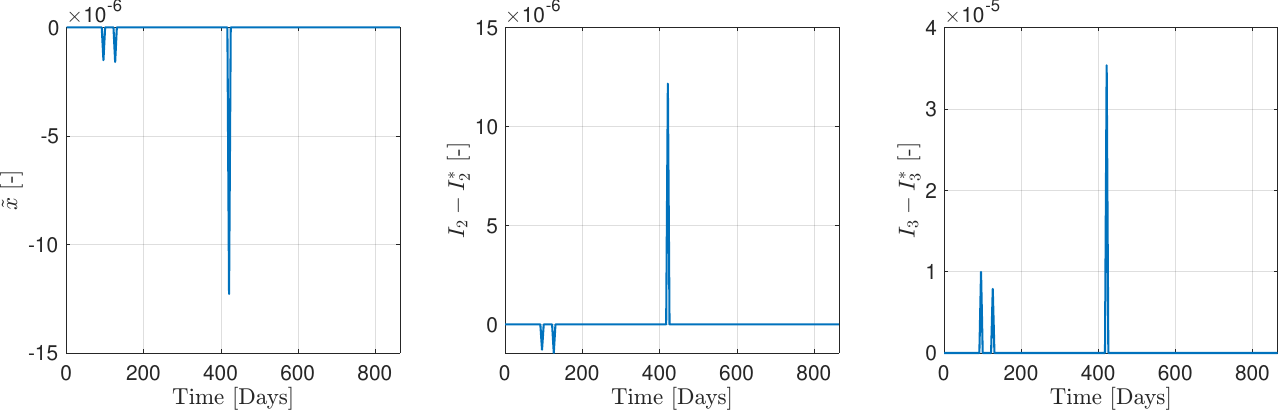}
    \caption{Elements of $\mathbf{F}(\mathbf{v})$ immediately after each impulsive maneuver.\label{fig:BirkhoffLissajous-AA}}
\end{figure}

Consider a Lissajous torus  near the EM $L_1$ libration point with actions $(I_2^*,I_3^*) = (0.1,0.007)$. An initial location on the torus, defined by $(\phi_2,\phi_3) = (0.25,0.1)$, is selected, along with a final time of $t_f=200\, TU$ and an increment of $\Delta t=1\,TU$. After propagating the restricted three-body state for $\Delta t$, \eqref{Birkhoffnewton} is iterated to obtain the optimal maneuver, $\Delta \mathbf{v}$. This sequence is then repeated until $t=t_f$. The results of this station-keeping scenario are shown in Fig. \ref{fig:BirkhoffLissajous} and \ref{fig:BirkhoffLissajous-AA}.

The controlled trajectory as well as the burn locations are plotted in Fig. \ref{subfig:BirkhoffLissajous-RTB}. There are a few locations along the trajectory at which the iterative process does not converge, which are indicated by the significant spikes in the $\tilde{x}$ coordinate shown in Fig. \ref{fig:BirkhoffLissajous-AA}. At these points, the Jacobian in \eqref{jac} is nearly singular, and the iterative process is artificially ended at 20 iterations, with the output being the current $\Delta \mathbf{v}$ at the final iteration. The average cost of station-keeping this specific trajectory utilizing the Birkhoff formulation is $\|\Delta \mathbf{v}\|=7.357\,m/s$ per year.

\newpage
\subsection{Resonant Station-keeping}
\subsubsection{Non-halo Formulation}\hfill

The station-keeping of a resonant normal form Lissajous or quasihalo trajectory requires the definition of a slightly different error vector than the one used in the Birkhoff case. The action $\hat{I}_3$ is equal to the sum of the old $I_2$ and $I_2$ actions, meaning that by maintaining the desired $\hat{I}_3$ of a given trajectory, it is hypothesized that the fluctuation of the $\hat{I}_2$ of the controlled trajectory will remain close to the intended time history of $\hat{I}_2$ obtained by propagating $\mathbf{x}_{AA}^R$ in the action-angle space. The explicit presence of $\hat{I}_2$ in the error vector is unnecessary, resulting in the following error vector, $\mathbf{F}$ for the station-keeping of quasiperiodic trajectories leveraging the resonant normal form.
\begin{equation}\label{eq:errorvec2}
    \mathbf{F}(\mathbf{v}) =\begin{bmatrix}\tilde{x}\\ \hat{I}_3-\hat{I}_3^* \end{bmatrix} 
\end{equation}
To account for the lower dimension of $\mathbf{F}(\mathbf{v})$, the iterative scheme defined in \eqref{Birkhoffnewton} must be altered slightly to account for the non-square nature of the new Jacobian matrix.
\begin{equation}\label{Resonantnewton}
    \mathbf{v}_{k+1}=\mathbf{v}_k - \left(\left.\frac{d \mathbf{F}}{d \mathbf{v}}\right|_{\mathbf{v}_k}\right)^{\dagger}\mathbf{F}(\mathbf{v}_k)
\end{equation}
The $\dagger$ notation denotes the pseudoinverse for an underdetermined system: $H^\dagger = H^T\left(HH^T\right)^{-1}$. The Jacobian in \eqref{Resonantnewton} will simply be the following.
\begin{equation}
    \frac{d \mathbf{F}}{d \mathbf{v}} =\left[\begin{matrix}\frac{d \tilde{x}}{d \dot{x}}& \frac{d \tilde{x}}{d \dot{y}}& \frac{d \tilde{x}}{d \dot{z}}\\ \frac{d \hat{I}_3}{d \dot{x}}& \frac{d\hat{I}_3}{d \dot{y}}& \frac{d \hat{I}_3}{d \dot{z}} \end{matrix}\right] 
\end{equation}
Fortunately, no new derivatives need to be calculated, because
\begin{equation}
    \frac{d \hat{I}_3}{d \mathbf{x}_{RTB}} = \frac{d I_2}{d \mathbf{x}_{RTB}} + \frac{d I_3}{d \mathbf{x}_{RTB}},
\end{equation}
where both components have already been derived in the previous section. Clearly the resonant station-keeping formulation for non-halo trajectories is just a slight augmentation of the Birkhoff case.

\subsubsection{Non-halo Results}\hfill

For the sake of comparison with the results obtained from the Birkhoff station-keeping scenario, an extremely similar Earth-Moon $L_1$ Lissajous torus is first considered. The parameters of the scenario are: $(\hat{I}_2(0),\hat{I}_3^*)=(0.1,0.107)$, $\Delta t=1\,TU$, and $t_f=200\,TU$.

\begin{figure}[htb!]
    \centering
    \subfigure[\label{subfig:ResonantLissajous-RTB}]{
        {\includegraphics[width=0.6\textwidth]{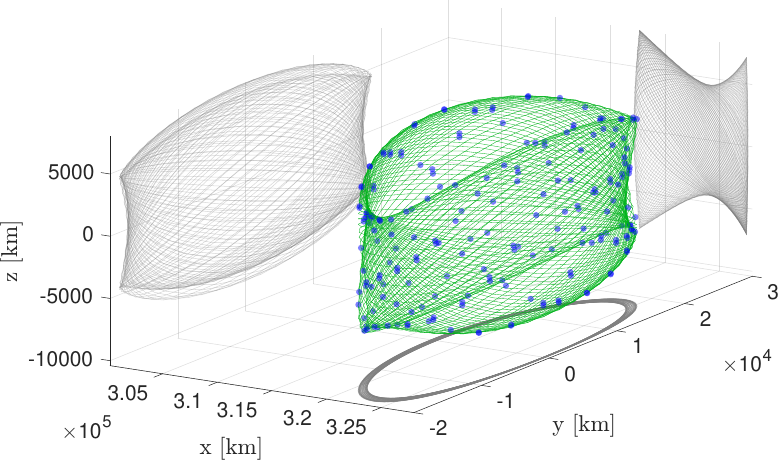}}
        }
    \subfigure[\label{subfig:ResonantLissajous-Control}]{
        {\includegraphics[width=0.35\textwidth]{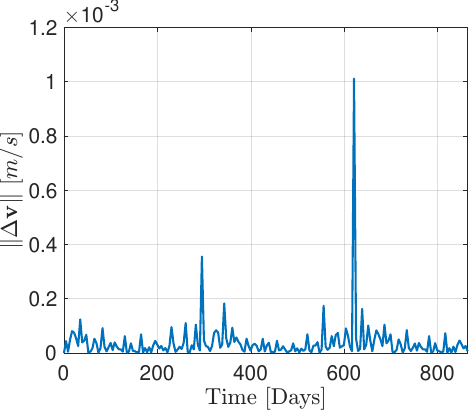}}
        }
        \caption{Station-keeping of an $L_1$ resonant Lissajous trajectory ($t_f = 200\,TU$,\,\,$\Delta t = 1\,TU$).\label{fig:BirkhoffResonant}}
\end{figure}
\begin{figure}[htb!]
    \centering
    \includegraphics[width=1\textwidth]{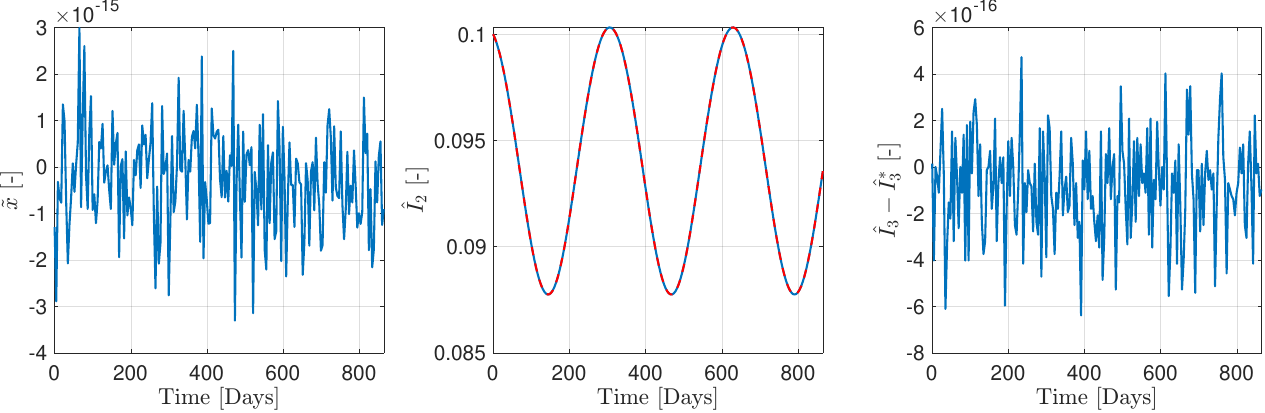}
    \caption{Elements of $\mathbf{F}(\mathbf{v})$ immediately after each impulsive maneuver, as well as a comparison between the controlled and intended time histories of $\hat{I}_2$.\label{fig:ResonantLissajous-AA}}
\end{figure}
It is clear from Fig. \ref{subfig:ResonantLissajous-Control} that the control input required to stay on the resonant Lissajous torus is several orders of magnitude lower than that which was required for the station-keeping of the Birkhoff Lissajous trajectory. The average cost of station-keeping this particular trajectory is $\|\Delta\mathbf{v}\|=0.0032\,m/s$ per year, with a third of the total cost resulting from one outlier that occurs at $t=144\,TU$ ($\approx$ 635 days).

The center graph in Fig. \ref{fig:ResonantLissajous-AA} confirms the hypothesis that by explicitly driving the error in $\hat{I}_3$ to zero (in addition to $\tilde{x}$), the proper fluctuation in $\hat{I}_2$ is maintained. The $\hat{I}_2$ of the controlled trajectory is plotted in blue, while the anticipated time history of $\hat{I}_2$ is plotted as a dashed red curve. The two plots overlap completely for the full duration of the station-keeping scenario. This means that not only is the spacecraft kept on the correct Lissajous torus, but it is also kept on the correct Lissajous \textit{trajectory}, meaning that there is no phase error accumulating between the spacecraft's controlled state and its planned state obtained initially through propagation in the action-angle space.

While the results are promising for the noiseless scenario, it is important to estimate how the station-keeping approach will be affected by the introduction of errors. Inspired by \cite{folta2010stationkeeping}, an error in maneuver impulse magnitude is introduced by scaling the optimal $\Delta \mathbf{v}$ by a certain percentage before applying every maneuver, i.e., $\Delta\mathbf{v} = (1+p)\Delta\mathbf{v}^*$. Results are included in Table \ref{tab:controlerror}, and show that while the approach is fairly robust to small errors, the cost of station-keeping quickly increases for greater errors. The spacecraft departs from the Lissajous torus at around $t=30\, TU$ when a $10\%$ maneuver error is applied.
\begin{table}[htb!]
\centering
\begin{tabular}{ c|c } 
\centering
Maneuver Error & $\|\Delta\mathbf{v}\|$ (per year) \\
\hline
$0\%$ & 0.00318 $m/s$\\ 
$1\%$ & 0.00335 $m/s$\\ 
$5\%$ & 0.01902 $m/s$\\ 
\end{tabular}
\caption{Average Lissajous station-keeping costs for different errors in impulse maneuver magnitude.\label{tab:controlerror}}
\end{table}

The lower cost associated with the resonant station-keeping is another indication of the advantages of the resonant normal form over the Birkhoff normal form. Assuming perfect station-keeping (no numerical errors, singular Jacobians, etc.), the lower bound on $\|\Delta\mathbf{v}\|$ will be proportional to the difference between the true numerical desired trajectory and its normal form approximation. Figures \ref{subfig:BirkhoffLissajous-Control} and \ref{subfig:ResonantLissajous-Control} make a strong case in favor of using the resonant normal form for station-keeping purposes.

Moving on from Lissajous trajectories, the same station-keeping approach defined in \eqref{eq:errorvec2} and \eqref{Resonantnewton} can be applied directly to quasihalo trajectories. Consider a northern Earth-Moon $L_1$ quasihalo torus with $(\hat{I}_2(0),\hat{I}_3^*)=(0.225,0.258)$, and initial location $(\theta_2(0),\theta_3(0)) = (\frac{\pi}{2},0)\, rad$. 
\begin{figure}[htb!]
    \centering
    \subfigure[\label{subfig:LargeQuasiHalo-RTB}]{
        {\includegraphics[width=0.45\textwidth]{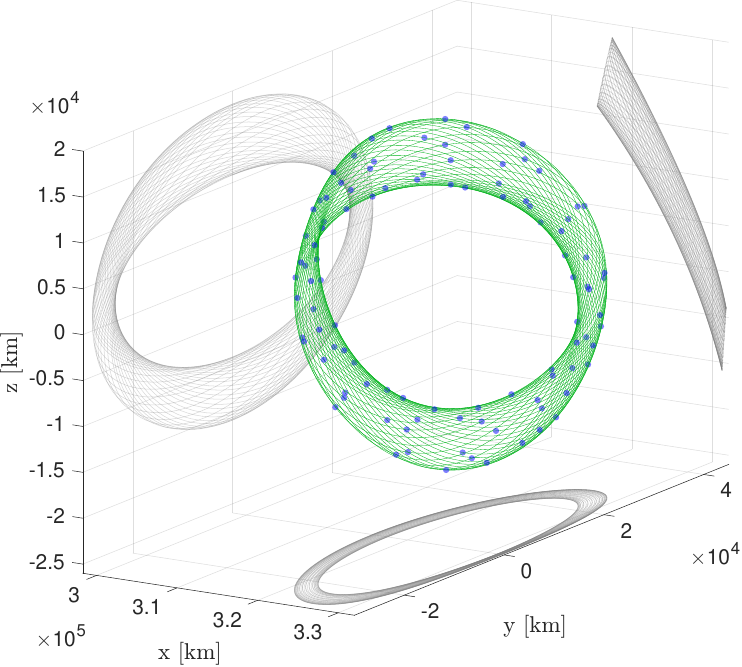}}
        }
    \subfigure[\label{subfig:LargeQuasiHalo-Control}]{
        {\includegraphics[width=0.4\textwidth]{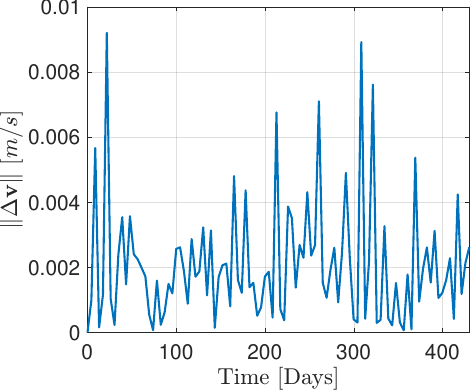}}
        }
        \caption{Station-keeping of an $L_1$ quasihalo  trajectory ($t_f = 100\,TU$,\,\,$\Delta t = 1\,TU$).\label{fig:LargeQuasiHalo}}
\end{figure}
\begin{figure}[htb!]
    \centering
    \includegraphics[width=1\textwidth]{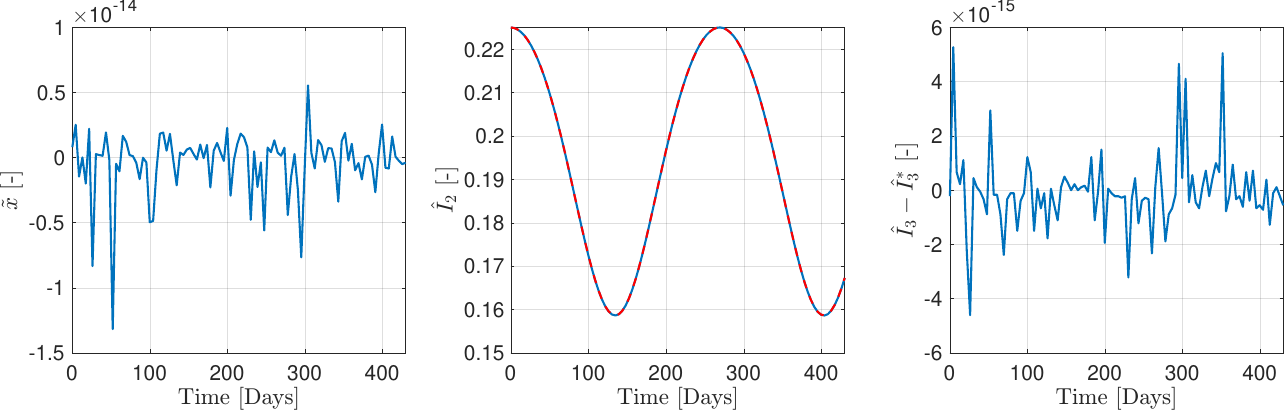}
    \caption{Elements of $\mathbf{F}(\mathbf{v})$ immediately after each impulsive maneuver, as well as a comparison between the controlled and intended time histories of $\hat{I}_2$.\label{fig:LargeQuasiHalo-AA}}
\end{figure}
Just like in the case of the Lissajous trajectory, Fig. \ref{fig:LargeQuasiHalo} and \ref{fig:LargeQuasiHalo-AA} show that the resonant station-keeping approach defined in \eqref{eq:errorvec2} and \eqref{Resonantnewton} performs well for quasihalo trajectories. The higher cost is fairly expected, due to the normal form's ability to better approximate Lissajous trajectories. Unlike the previous scenarios, there are no maneuvers that could be considered significant outliers. This could be a result of the smoother shape of the quasihalo torus, which reduces the possibility of encountering a singular Jacobian.

Additional analysis was performed in which the interval between maneuvers, $\Delta t$, was increased slightly. Intuitively, a larger $\Delta t$ will result in a greater overall cost, since a larger $\tilde{x}$ component will need to be zeroed out at each burn location. Table \ref{tab:interval} shows that the cost increases by an order of magnitude as the interval is increased by $1\,TU$. This trend is unsustainable, as for too large a value of $\Delta t$, the spacecraft will depart from the neighborhood of the desired trajectory, making the Newton iteration scheme in \eqref{Resonantnewton} invalid. 
\begin{table}[htb!]
\centering
\begin{tabular}{ c|c } 
\centering
$\Delta t$ & $\|\Delta\mathbf{v}\|$ (per year) \\
\hline
$1 \, TU$ & 0.1773 $m/s$\\ 
$2 \, TU$ & 1.4695  $m/s$\\ 
$3 \, TU$ & 13.001  $m/s$\\ 
\end{tabular}
\caption{Average station-keeping cost of the quasihalo trajectory for different values of $\Delta t$.\label{tab:interval}}
\end{table}

This section will not show any results for the application of the resonant station-keeping approach to Lyapunov or vertical orbits, as they are considered more trivial cases, however, the same process can be effectively applied to these periodic orbits. The only orbit family (that the resonant normal form is able to approximate) to which the approach outlined in \eqref{eq:errorvec2} and \eqref{Resonantnewton} cannot be applied is the halo family. Indeed, a slightly different methodology is required for the station-keeping of halo orbits.

\subsubsection{Halo Formulation}\hfill

If the station-keeping formulation described in \eqref{eq:errorvec2} and \eqref{Resonantnewton} is supplied with an initial condition on a halo orbit, the resulting controlled trajectory will end up on a quasihalo torus with the same $\hat{I}_3$. Therefore, an additional constraint is required for the station-keeping of halo orbits. There are two candidates for the additional constraint, $\hat{I}_2$ and $\theta_2$, as both are constant along a halo orbit. Choosing $\theta_2$ as the additional constraint is the better option of the two, with the rationale being that $\hat{I}_2 = I_2$ and $\hat{I}_3 = I_2+I_3$, meaning that there is a greater likelihood that the Jacobian would not be full rank at particular locations along the orbit. 

Rather than defining an explicit constraint on the value of $\theta_2$, such as $\theta_2-\theta_2^*,\,\text{with}\,\theta_2^*=\pm\frac{\pi}{2}$, the fact that $\cos(\pm\frac{\pi}{2})=0$ will be taken advantage of by introducing $\gamma:=\cos(\theta_2)$ as the new constraint.
\begin{equation}\label{eq:errorvec3}
    \mathbf{F}(\mathbf{v}) = \begin{bmatrix}\tilde{x} \\ \gamma \\ \hat{I}_3-\hat{I}_3^* \end{bmatrix}
\end{equation}
The derivative of the error vector with respect to the velocity in the restricted three-body frame is a $3\times 3$ matrix,
\begin{equation}\label{eq:jac3}
    \frac{d \mathbf{F}}{d \mathbf{v}} =\left[\begin{matrix}\frac{d \tilde{x}}{d \dot{x}}& \frac{d \tilde{x}}{d \dot{y}}& \frac{d \tilde{x}}{d \dot{z}}\\ \frac{d \gamma}{d \dot{x}}& \frac{d\gamma}{d \dot{y}}& \frac{d \gamma}{d \dot{z}}\\ \frac{d \hat{I}_3}{d \dot{x}}& \frac{d\hat{I}_3}{d \dot{y}}& \frac{d \hat{I}_3}{d \dot{z}} \end{matrix}\right], 
\end{equation}
meaning that the Newton iteration for the station-keeping of a halo orbit will be identical to \eqref{Birkhoffnewton}. The only new derivatives that must be derived are the ones that appear in the second row of \eqref{eq:jac3}. The derivative of $\gamma$ with respect to $\mathbf{x}_{RTB}$ can be expanded as
\begin{equation}
    \frac{d \gamma}{d \mathbf{x}_{RTB}} = \frac{d \gamma}{d \mathbf{x}_{NF}}\frac{d \mathbf{x}_{NF}}{d \mathbf{x}_{RTB}},
\end{equation}
where the derivative of $\mathbf{x}_{NF}$ with respect to $\mathbf{x}_{RTB}$ is already known, and the remaining derivative is given by
\begin{equation}\label{dgammadnf}
    \frac{d \gamma}{d \mathbf{x}_{NF}} = \begin{bmatrix}0&\frac{d \gamma}{d \tilde{y}}& \frac{d \gamma}{d \tilde{z}}& 0&\frac{d \gamma}{d \tilde{p}_y}& \frac{d \gamma}{d \tilde{p}_z}  \end{bmatrix}.
\end{equation}
Let $\boldsymbol{\xi}:=[\tilde{y},\tilde{p}_y]^T$ and $\boldsymbol{\eta}:=[\tilde{z},\tilde{p}_z]^T$, then
\begin{equation}
    \gamma = \frac{\boldsymbol{\xi}\cdot\boldsymbol{\eta}}{\|\boldsymbol{\xi}\|\|\boldsymbol{\eta}\|} = \frac{\tilde{y}\tilde{z} + \tilde{p}_y\tilde{p}_z}{\sqrt{\tilde{y}^2+\tilde{p}_y^2+\tilde{z}^2+\tilde{p}_z^2}}.
\end{equation}
With $\gamma$ written explicitly in terms of the normal form coordinates, it is straightforward to find the derivatives in \eqref{dgammadnf}.
\begin{align}
    \frac{d \gamma}{d \tilde{y}} &= \frac{\tilde{z}\|\boldsymbol{\xi}\|^2\|\boldsymbol{\eta}\|^2 - \tilde{y}\left(\boldsymbol{\xi}\cdot\boldsymbol{\eta}\right)}{\|\boldsymbol{\xi}\|^3\|\boldsymbol{\eta}\|^3} \\ 
    \frac{d \gamma}{d \tilde{z}} &= \frac{\tilde{y}\|\boldsymbol{\xi}\|^2\|\boldsymbol{\eta}\|^2 - \tilde{z}\left(\boldsymbol{\xi}\cdot\boldsymbol{\eta}\right)}{\|\boldsymbol{\xi}\|^3\|\boldsymbol{\eta}\|^3} \\ 
    \frac{d \gamma}{d \tilde{p}_y} &= \frac{\tilde{p}_z\|\boldsymbol{\xi}\|^2\|\boldsymbol{\eta}\|^2 - \tilde{p}_y\left(\boldsymbol{\xi}\cdot\boldsymbol{\eta}\right)}{\|\boldsymbol{\xi}\|^3\|\boldsymbol{\eta}\|^3} \\ 
    \frac{d \gamma}{d \tilde{p}_z} &= \frac{\tilde{p}_y\|\boldsymbol{\xi}\|^2\|\boldsymbol{\eta}\|^2 - \tilde{p}_z\left(\boldsymbol{\xi}\cdot\boldsymbol{\eta}\right)}{\|\boldsymbol{\xi}\|^3\|\boldsymbol{\eta}\|^3}  
\end{align}
Once again, a station-keeping approach is defined by solving \eqref{eq:errorvec3} and \eqref{Birkhoffnewton} for the optimal $\Delta\mathbf{v}$, and then performing the impulsive burn and propagating the spacecraft's state for a specified interval of $\Delta t$. This sequence is repeated until a final time of $t_f$ is reached.

\subsubsection{Halo Results}\hfill

\begin{figure}[htb!]
    \centering
    \subfigure[\label{subfig:HaloImpulseLong-RTB}]{
        {\includegraphics[width=0.5\textwidth]{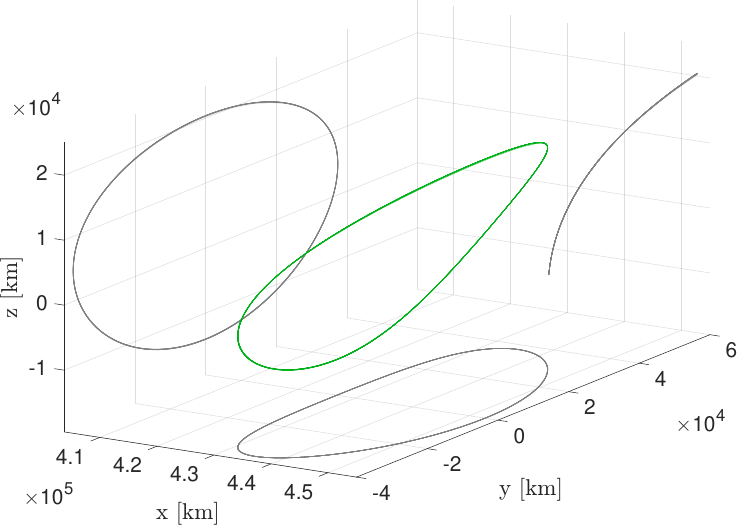}}
        }
    \subfigure[\label{subfig:HaloImpulseLong-Control}]{
        {\includegraphics[width=0.4\textwidth]{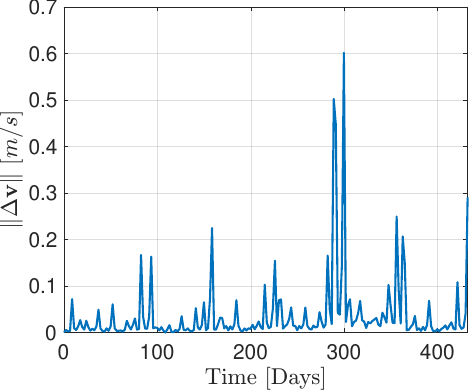}}
        }
        \caption{Station-keeping of a northern $L_2$ halo orbit ($t_f = 100\,TU$,\,\,$\Delta t = 0.5\,TU$).\label{fig:HaloImpulseLong}}
\end{figure}

\begin{figure}[htb!]
    \centering
    \includegraphics[width=1\textwidth]{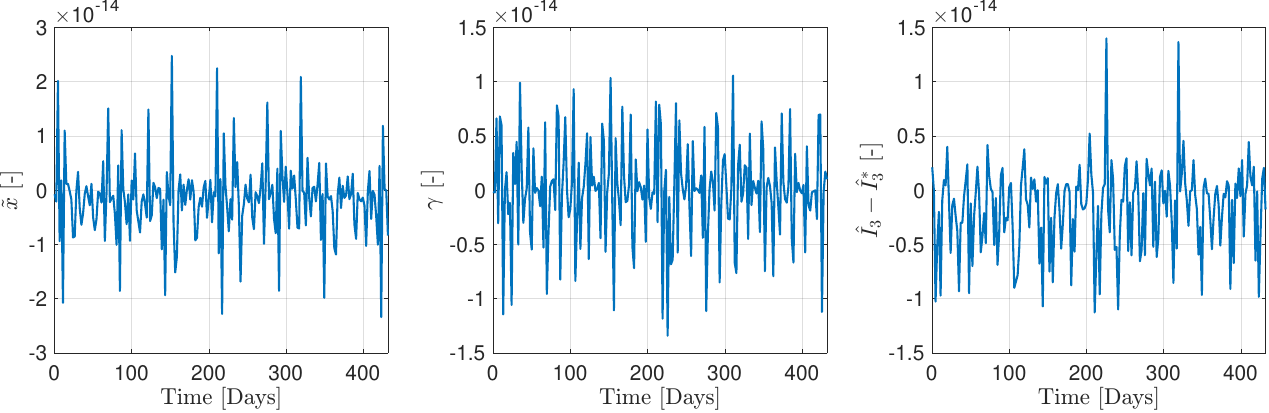}
    \caption{Elements of $\mathbf{F}(\mathbf{v})$ immediately after each impulsive maneuver.\label{fig:HaloImpulseLong-AA}}
\end{figure}

Consider a northern Earth-Moon halo orbit near the $L_2$ libration point, with actions given by $(\hat{I}_2,\hat{I_3})\approx(0.2431,0.3076)$ and initial location $(\theta_2,\theta_3(0)) = (-\frac{\pi}{2},0.1)\,rad$. Figures \ref{fig:HaloImpulseLong} and \ref{fig:HaloImpulseLong-AA} show the results of station-keeping this particular halo orbit with maneuver spacing of $\Delta t=0.5\,TU$ and a total duration of $t_f=100\,TU$.

The burn locations have not been included in Fig. \ref{subfig:HaloImpulseLong-RTB} as they densely populate and obscure the controlled trajectory. The average cost of this station-keeping scenario is $\|\Delta\mathbf{v}\|=6.050\,m/s$ per year. At one particular time, $t=25\,TU$ ($\approx$ 108.6 days), the Newton iteration is unable to converge to a value of $\tilde{x}$ below $1e-6$, and so no $\Delta \mathbf{v}$ is applied. The iterative scheme is able to converge at all other 199 maneuver times. Due to the fact that proper convergence is achieved at the maneuver time immediately following the bad case, it can be inferred that the issue stems from particular points on the orbit at which the Jacobian is nearly singular.
 If the same station-keeping scenario is considered with $\Delta t=1\,TU$, the average cost increases to $8.654\,m/s$ per year, and there is slight (but visible) separation at the top of the halo orbit.
\newpage
\section{Conclusion}
This paper presented the means for characterizing several families of trajectories near the librations point in the CR3BP and provided a comparison between the Birkhoff and Resonant normal forms. It was found that the additional terms retained in the resonant normal form substantially increase the approximation accuracy of out-of-plane motion. This increased region of validity enhances the approximation of the vertical family, the halo family, and Lissajous trajectories.  Not only that, but it allows for the parameterization of quasihalo trajectories-- a feat that the Birkhoff normal form cannot replicate. It was also shown that the resonant normal form provides sufficiently accurate quasiperiodic tori, which can be used as the initial guess in a Poincar\'e section approach to determining their numerical equivalents. This eliminates the need to start from some base halo orbit, find a small quasihalo using the monodromy matrix, and then continue outward in a parameter (such as cross-sectional area) until the desired numerical torus is reached, which is the most commonly used method.

Additionally, a method of station-keeping analogous to Floquet modes was formulated which leverages the simple normal form parameterization of both periodic and quasiperiodic trajectories. Initial results are promising, however, more analysis is required to study the effects of additional sources of error, such as tracking error and maneuver pointing error. The main drawback of this approach to station-keeping is that the average cost in terms of $\|\Delta\mathbf{v}$ will increase as the distance from the libration point increases. 

Future work will include introducing a variable maneuver interval $\Delta t$, where the $\tilde{x}$ component will be monitored and activate a maneuver upon reaching a threshold value. To avoid specific locations on a trajectory that produce convergence issues for the Newton iteration, a temporary control scheme could be introduced that avoids the use of near-singular Jacobians, or even simple keep-out zones could be implemented if needed.
\section{Acknowledgments}
This material is based upon work supported jointly by the AFOSR grant FA9550-23-1-0512 and the Penn State Applied Research Lab (ARL) Walker Assistantship program.

\bibliographystyle{AAS_publication}
\bibliography{references}

\begin{thebibliography}{10}

\bibitem{cellettiLissajousHaloOrbits2015}
A.~Celletti, G.~Pucacco, and D.~Stella, ``Lissajous and {{Halo Orbits}} in the
  {{Restricted Three-Body Problem}},''  {\em Journal of Nonlinear Science},
  Vol.~25, Apr. 2015, pp.~343--370, 10.1007/s00332-015-9232-2.

\bibitem{peterson2024toolkit}
L.~T. Peterson and D.~J. Scheeres, ``Local Orbital Element Toolkit in Cislunar
  Space,''  {\em 2024 AAS/AIAA Astrodynamics Specialist Conference}, 2024.

\bibitem{farres2022geometrical}
A.~Farr{\'e}s, C.~Gao, J.~J. Masdemont, G.~G{\'o}mez, D.~C. Folta, and
  C.~Webster, ``Geometrical analysis of station-keeping strategies about
  libration point orbits,''  {\em Journal of Guidance, Control, and Dynamics},
  Vol.~45, No.~6, 2022, pp.~1108--1125.

\bibitem{gomezStationKeepingStrategiesTranslunar1998}
G.~Gomez, K.~Howell, J.~Masdemont, and C.~Simo, ``Station-{{Keeping Strategies
  For Translunar Libration Point Orbits}},''  {\em Advances in Astronautical
  Sciences}, Vol.~94, Jan. 1998.

\bibitem{schwabCislunarTransportCharacterization2024}
D.~Schwab, {\em Cislunar {{Transport Characterization}} for {{Space Situational
  Awareness}}}.
\newblock PhD thesis, The Pennsylvania State University, University Park, PA,
  May 2024.

\bibitem{jorbaDynamicsCenterManifold1999}
{\'A}.~Jorba and J.~Masdemont, ``Dynamics in the Center Manifold of the
  Collinear Points of the Restricted Three Body Problem,''  {\em Physica D:
  Nonlinear Phenomena}, Vol.~132, July 1999, pp.~189--213,
  10.1016/S0167-2789(99)00042-1.

\bibitem{jorbaMethodologyNumericalComputation1999}
{\'A}.~Jorba, ``A {{Methodology}} for the {{Numerical Computation}} of {{Normal
  Forms}}, {{Centre Manifolds}} and {{First Integrals}} of {{Hamiltonian
  Systems}},''  {\em Experimental Mathematics}, Vol.~8, Jan. 1999,
  pp.~155--195, 10.1080/10586458.1999.10504397.

\bibitem{schwabCharacterizingAccuracyNormal2024}
D.~Schwab, R.~Eapen, and P.~Singla, ``Characterizing {{Accuracy}} of {{Normal
  Forms}} to {{Study Trajectories}} in {{Cislunar Space}},''  {\em The Journal
  of the Astronautical Sciences}, Vol.~71, Mar. 2024, p.~16,
  10.1007/s40295-024-00440-z.

\bibitem{petersonOrbitalElementsRestricted2022}
L.~T. Peterson and D.~J. Scheeres, ``Orbital {{Elements}} for the {{Restricted
  Three-Body Problem}},''  {\em 2022 {{AAS}}/{{AIAA Astrodynamics Specialist
  Conference}}}, Charlotte, North Carolina, Aug. 2022, p.~20.

\bibitem{jorbaNumericalComputationNormal1998}
{\'A}.~Jorba and J.~Villanueva, ``Numerical Computation of Normal Forms around
  Some Periodic Orbits of the Restricted Three-Body Problem,''  {\em Physica D:
  Nonlinear Phenomena}, Vol.~114, Apr. 1998, pp.~197--229,
  10.1016/S0167-2789(97)00194-2.

\bibitem{angeljorbaLagrangianSolutions2015}
{\'A}.~Jorba, ``The {{Lagrangian Solutions}},''  {\em UNESCO Encyclopedia of
  Life Support Systems}, Vol.~6.119.55, Jan. 2015.

\bibitem{zhaoLieseriesTransformationsApplications2023b}
S.~Zhao and H.~Lei, ``Lie-Series Transformations and Applications to
  Construction of Analytical Solutions,''  Feb. 2023,
  10.21203/rs.3.rs-2523599/v1.

\bibitem{kolemen2012multiple}
E.~Kolemen, N.~J. Kasdin, and P.~Gurfil, ``Multiple Poincar{\'e} sections
  method for finding the quasiperiodic orbits of the restricted three body
  problem,''  {\em Celestial Mechanics and Dynamical Astronomy}, Vol.~112,
  2012, pp.~47--74.

\bibitem{folta2010stationkeeping}
D.~Folta, T.~Pavlak, K.~Howell, M.~Woodard, and D.~Woodfork, ``Stationkeeping
  of Lissajous trajectories in the Earth-Moon system with applications to
  ARTEMIS,''  {\em AAS/AIAA Space Flight Mechanics Meeting},
  No.~LEGNEW-OLDGSFC-GSFC-LN-1055, 2010.

\end{thebibliography}

\end{document}